\documentclass[twocolumn,aps,prb,amsfonts,amssymb,amsmath,showpacs,english]{revtex4}
\usepackage[latin9]{inputenc}
\usepackage{amssymb}
\usepackage{esint}
\usepackage{babel}
\usepackage{graphicx}
\usepackage{mathrsfs}
\usepackage{dcolumn}% Align table columns on decimal point
\usepackage{bm}% bold math
\usepackage{color}% color

\begin{document}

%\preprint{APS/123-QED}

\title{Surface states scattering from a step defect in topological insulator $\mathrm{Bi}_{2}\mathrm{Te}_{3}$ }

\author{Jin An$^{1,2}$ and C. S. Ting$^1$}

\affiliation{$^{1}$Texas Center for Superconductivity and Department of Physics, University of Houston, Houston,
Texas 77204, USA \\
$^{2}$National Laboratory of Solid State Microstructures and Department of Physics, Nanjing University, Nanjing 210093, China}

\date{\today}

\begin{abstract}
We study theoretically  the general scattering problem of a straight step defect on the surface of topological insulator $\mathrm{Bi}_{2}\mathrm{Te}_{3}$ with strong warping effect using a quantum-mechanical approach. At high energy where the warping effect is large, an incident electron on a step defect running along $\Gamma-\mathrm{M}$ may exhibit perfect transmission whereas on a defect running along $\Gamma-\mathrm{K}$ has a finite probability to be reflected and may even exhibit resonant total reflection. The transmission property in the latter case is also sensitive to whether there is  particle-hole symmetry in the system. Although backscattering is prohibited by time reversal symmetry, the nearly normal incident electron on a defect running along $\Gamma-\mathrm{K}$ has a finite reflection. This is interpreted as the consequence of the existence of decaying modes localized at the defect, acting as a magnetic barrier for the propagating modes. The predicted Friedel oscillations and the power-law decaying behavior of the local density of states (LDOS) near the defect are in good agreement with recent scanning tunneling microscopy experiments on $\mathrm{Bi}_{2}\mathrm{Te}_{3}$. The high-energy LDOS of the surface states is also found to show the multi-periodic Friedel oscillations, caused by competing characteristic scattering processes.

\end{abstract}

% PACS, the Physics and Astronomy
                             % Classification Scheme.
%\keywords{Suggested keywords}%Use show keys class option if keyword
                              %display desired

\maketitle
\section{introduction}
Since the prediction of three dimensional (3D) strong topological insulator (TI),\cite{LiangFu} which is characterized by a bulk gap and an odd number of gapless Dirac cones on the surface,\cite{LiangFu,XLQi,XLQi2} it has been found by angle-resolved-photoemission spectroscopy(ARPES) to be realized in $\mathrm{Bi}_{x}\mathrm{Sb}_{1-x}$,\cite{Hsieh,Hsieh2,Roushan} $\mathrm{Bi}_{2}\mathrm{Se}_{3}$,\cite{Xia,Hsieh4} and $\mathrm{Bi}_{2}\mathrm{Te}_{3}$\cite{YLChen,Hsieh3} materials. The surface states of a TI are two-dimensional (2D) helical Dirac fermions, topologically protected by time reversal symmetry. These fermions dominate the quantum transport of a TI, and may give rise to potential applications in the future electronic devices.\cite{XLQi} In the presence of imperfections on the surface of a TI, the interference patterns of electrons can be probed by scanning tunneling microscopy (STM), which directly measures the LDOS near the imperfections. Since surface electrons are prevented from backscattering due to time-reversal symmetry, it is more difficult for them to be scattered, compared with electrons in conventional metals. For 2D Dirac electrons, the Friedel oscillations of the LDOS near a nonmagnetic impurity or a line defect are predicted to show a $r^{-2}$,\cite{WangQH,QinLiu} or $r^{-3/2}$,\cite{WangQH,QinLiu,HuJP,Balatsky,Biswas} asymptotic power-law decay with $r$  as the distance from the imperfection, consistent with recent STM experiments on TIs\cite{ZhuBF}. Although the imperfection scattering effect on the surface is well understood for pure 2D Dirac fermions, both in terms of TIs and graphene,\cite{Katsnelson} much less is known for the case with a  distorted Dirac cone caused by the crystal symmetry of TI materials.\cite{Xia,YLChen,Hsieh3} Among the discovered 3D TIs, $\mathrm{Bi}_{2}\mathrm{Te}_{3}$ is found by ARPES to have a strong warping effect,\cite{YLChen,LFu,Hsieh3} while surface electronic structure probed by STM experiments\cite{XueQK,STM,ZhuBF,Alpichshev,Alpichshev2} is still not conclusive.  Therefore further theoretical and experimental studies are needed.

In this paper, based on a quantum-mechanical approach we explore  the electronic structure of the surface states in a TI with a strong warping effect, under the influence of a straight atomic step defect, or edge defect, which is modeled by us as a one-dimensional delta function.\cite{ZhangDG} We find that at higher energy where the warping effect is large, there are several critical $\textbf{k}$ points on the equal-energy contour. An incoming electron with these momenta will be totally reflected or perfectly transmitted. When the defect is along $\Gamma-\mathrm{K}$, the existence of localized decaying modes causes a finite reflection for nearly normal incident electrons. In another word, as long as the incident direction is not perfectly normal, finite reflection occurs.  The Friedel oscillations of high-energy LDOS are dominated by the competition between several scattering processes. Within a sufficient long distance (10nm-100nm) from the defect, the asymptotic Friedel oscillations of LDOS at low energies are found to be best fitted by a power-law decay function, while that at higher energies can be hardly fitted. The Fourier transform of the LDOS is predicted to show a generic broad peak at zero momentum, which can be seen as the observable feature of the existence of the decaying modes near the step defect.

The paper is organized as follows. We introduce our model and method in section II, where the scattering wave function and its corresponding boundary condition in the presence of warping effect are discussed in detail. In section III, we demonstrate the warping effect on the transmission property of an incident electron, and then give the physical explanation of the phenomenon of the finite reflection for nearly normal incidence when the defect is along $\Gamma-\mathrm{K}$. In section IV, we show the numerical LDOS for the two situations where the step defect runs along $\Gamma-\mathrm{M}$ and $\Gamma-\mathrm{K}$ respectively, and compare them with STM experiments. The LDOS is then Fourier transformed to further analyze the electrons' interference patterns near the step defect. In section V, we summarize our results.

\section{model and method}
For the energy range we focused on, we assume the interaction between the surface states and bulk ones can be totally neglected, so that the property of the surface states can be explored independently. ARPES experiments have revealed that $\mathrm{Bi}_{2}\mathrm{Te}_{3}$ has a single Dirac cone on its surface,\cite{YLChen} with distortion by a cubic spin-orbit coupling\cite{LFu} which still respects the three-fold symmetry of $\mathrm{Bi}_{2}\mathrm{Te}_{3}$ crystal. Therefore the single-particle Hamiltonian for the surface electrons of $\mathrm{Bi}_{2}\mathrm{Te}_{3}$ can be given by
\begin{eqnarray}
\nonumber H(p_{x},p_{y})=\frac{p^{2}}{2m^{*}}&+&v(p_x\sigma_y-p_y\sigma_x) \\
&+&\lambda p_x(p_x^{2}-3p_y^{2})\sigma_{z},
\end{eqnarray}%
where $p_{x}=-i\hbar\partial_{x}$, $p_{y}=-i\hbar\partial_{y}$ are the 2D momentum operators of surface electrons, and $\bm{\sigma}=(\sigma_{x},\sigma_{y},\sigma_{z})$ are the Pauli spin matrices.  $v$ is the Fermi velocity and $\lambda$ the warping parameter. The first term has been introduced to account for the possibility of particle-hole asymmetry. In unit of $\hbar=1$, the upper and lower bands in $k$ space are $E_{k}^{\pm}\equiv\frac{k^{2}}{2m^{*}}\pm E_\mathbf{k}$ with
\begin{eqnarray}
E_\mathbf{k}=\sqrt{v^{2}(k_{x}^{2}+k_{y}^{2})+\lambda^{2}k_{x}^{2}(k_{x}^{2}-3k_{y}^{2})^{2}}.
\end{eqnarray}%
For $\mathrm{Bi}_{2}\mathrm{Te}_{3}$, we choose $ \mathrm{v}=2.55\mathrm{eV}\cdot{{\mathrm{\AA}}}$, $\lambda=250\mathrm{eV}\cdot{\mathrm{\AA}}^{3}$, which produce the Fermi surface in good agreement with experiments.\cite{YLChen} Several representative equal-energy contours (EECs) in the absence of particle-hole asymmetry are shown in Fig.1, where the shapes of the curves change from a circle to hexagon and then to concave hexagrams.\cite{YLChen,LFu} Correspondingly the eigen-spinors for the upper and lower bands can be expressed as
$\chi_{\mathbf{k}}^{(\pm)}=\frac{1}{n_{\mathbf{k}}^{(\pm)}}(\begin{array}{c}
                                              \phi_\mathbf{k}\pm E_{\mathbf{k}} \\
                                              v(ik_{x}-k_{y})
                                            \end{array}
)$, with $\phi_{\mathbf{k}}=\lambda k_{x}(k_x^2-3k_y^2)$ and $n_\mathbf{k}^{(\pm)}$ the normalization factors. One immediate consequence of the warping effect on the spinors is that electron spin orientation $\mathbf{s}=<\bm{\sigma}>=\pm E_{k}^{-1}(-vk_{y},vk_{x},\phi_{\mathbf{k}})$ has an out-of-plane spin component\cite{WCLee} with alternating signs along angular direction, leading to the spin texture structure of the Dirac cone.\cite{SpinTexture}

\begin{figure}[!htb]
\includegraphics[width=0.5\textwidth,bb=0 0 150 105]{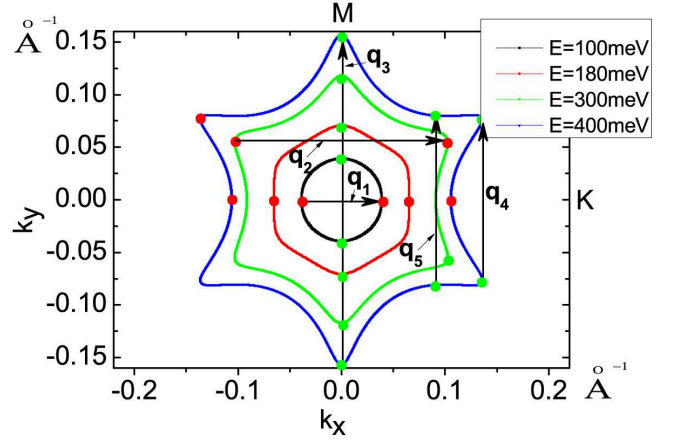}
\caption{\label{fig1} The equal-energy contours in the absence of particle-hole asymmetry for four different energies, which are measured referenced to the Dirac point. The red and green solid dots are the pairs of the extreme points which are responsible for the LDOS oscillations when the step defect is running along y ($\Gamma-M$) or x ($\Gamma-K$), respectively. Some characteristic scattering wave vectors are shown explicitly (solid arrows). }
\end{figure}

To explore the surface atomic step effect, we model the step as a delta-function scattering potential $U(\mathbf{r})$, rather than a step-function potential.\cite{ZhuBF,HuJP} This is because electrons on the upper and lower surfaces on both sides of  the step are expected to have the same surface potential and there is no potential difference across the step defect. Notice that model Hamiltonian (1) is only well defined at relatively lower energy. The problem with the step-function potential is that if the potential difference between the two surfaces is too large, the energy of one side of the defect will be beyond the well-defined lower energy region, causing the problematic conclusion. In this paper we focus on the two representative situations where the step extends along either $\Gamma-M$ or $\Gamma-K$ with $U(\mathbf{r})$ given by $U\delta(x)$ or $U\delta(y)$ respectively. The step scattering problem is treated quantum mechanically, since this approach has been proved to be very successful in graphene\cite{Katsnelson} in studying Dirac fermion scattering problems. We remark here that very recently this quantum-mechanical approach has been applied to study the warping effect of TI for step defect extending along $\Gamma-\mathrm{K}$,\cite{Kobayashi,Hungary} modeled by step-function potential.

\emph{Step defect along y}($\Gamma-\mathrm{M}$) \emph{direction}. An incident electron plane wave $\psi_\mathbf{k}(\mathbf{r})=\frac{1}{2\pi}e^{i\mathbf{k}\cdot\mathbf{r}}\chi_{\mathbf{k}}^{(+)}$ from one side with momentum $\mathbf{k}=(k_{x},k_{y})$ and energy $E(>0)$ will be reflected back to the same side or transmitted into the other side. Since $k_{y}$ is a good quantum number and energy E is conserved in the scattering process, naively, the reflected and transmitted waves will be characterized by $\mathbf{k}_{r}\equiv(-k_{x},k_{y})$ and $\textbf{k}$. Actually, for fixed $k_{y}$, the equation $\frac{k^{2}}{2m^{*}}\pm E_{\mathbf{k}}=E$ which determines the reflected and transmitted wave vectors $k_{x}$ is a sextic algebraic equation given by,
\begin{eqnarray}\nonumber
 k_{x}^{'6}&-&(6k_{y}^{'2}+E_{a}^{'2})k_{x}^{'4}+(9k_{y}^{'4}-2E_{a}^{'2}k_{y}^{'2}+2E_{a}^{'}E^{'}+1)k_{x}^{'2} \\
 &+&(-E_{a}^{'2}k_{y}^{'4}+(2E_{a}^{'}E^{'}+1)k_{y}^{'2}-E^{'2})=0,
\end{eqnarray}%
where the \emph{dimensionless} quantities $k_{x}^{'}=k_{x}a$, $k_{y}^{'}=k_{y}a$, $E^{'}=E/E^{*}$ and $E_{a}^{'}=E_{a}/E^{*}$, with the length scale $a=\sqrt{\lambda/v}$, energy scale $E^{*}=v/a$, and the particle-hole asymmetric characteristic energy $E_{a}=(2m^{*}a^{2})^{-1}$($a\approx1\texttt{nm}$, $E^{*}\approx260\texttt{meV}$ for $\mathrm{Bi}_{2}\mathrm{Te}_{3}$). This equation has six roots in total: $\pm k_{x,\alpha}$, $\alpha=1,2,3$. In the previous studies,\cite{ZhangDG,HuJP} only two real roots have been taken into account. To our knowledge, this six-root scattering mechanism has not been considered before to study the interference patterns of electrons for defect along $\Gamma-\mathrm{M}$, which will be concentrated on in this paper. For different $k_{y}$, the solutions for any scattering states can be further classified into two categories.

For type $\mathbf{I}$ solution, $k_{x,1}$, $k_{x,2}$, $k_{x,3}$ are all real numbers, as demonstrated in Fig.2(a) and Fig.2(b). This solution becomes possible when energy $E$ is above $E_{c1}$ ($E_{c1}=\sqrt{11}/3^{3/4}E^{*}\approx1.45E^{*}$ in the case of $E_{a}=0$), which is the critical value of energy whose EEC has inflection points fulfilling $(\partial k_{y}/\partial k_{x})_{E}=(\partial^{2}k_{y}/\partial k_{x}^{2})_{E}=0$. The solution indicates that an incoming electron wave from the left will have three reflected waves propagating towards the left, and three transmitted waves propagating towards the right (see Fig.2(c)) . Interestingly, both the transmitted and reflected waves have one hole-like propagating wave satisfying $k_{x}v_{x}(\mathbf{k})<0$, while the other twos are electron-like satisfying $k_{x}v_{x}(\mathbf{k})>0$  (see Fig.2(a)), with $v_{x}(\mathbf{k}) =(\partial E/\partial k_{x})_{k_{y}}$ the electron's group velocity along x axis. Here $k_{x,1}$, $k_{x,2}$, $k_{x,3}$ are so chosen that $v_{x}(\mathbf{k})>0$, i.e., $k_{x,1}>0$, $k_{x,2}<0$, $k_{x,3}>0$ (Fig.2(b)).
\begin{figure}[!htb]
\includegraphics[width=0.47\textwidth,bb=0 0 130 225]{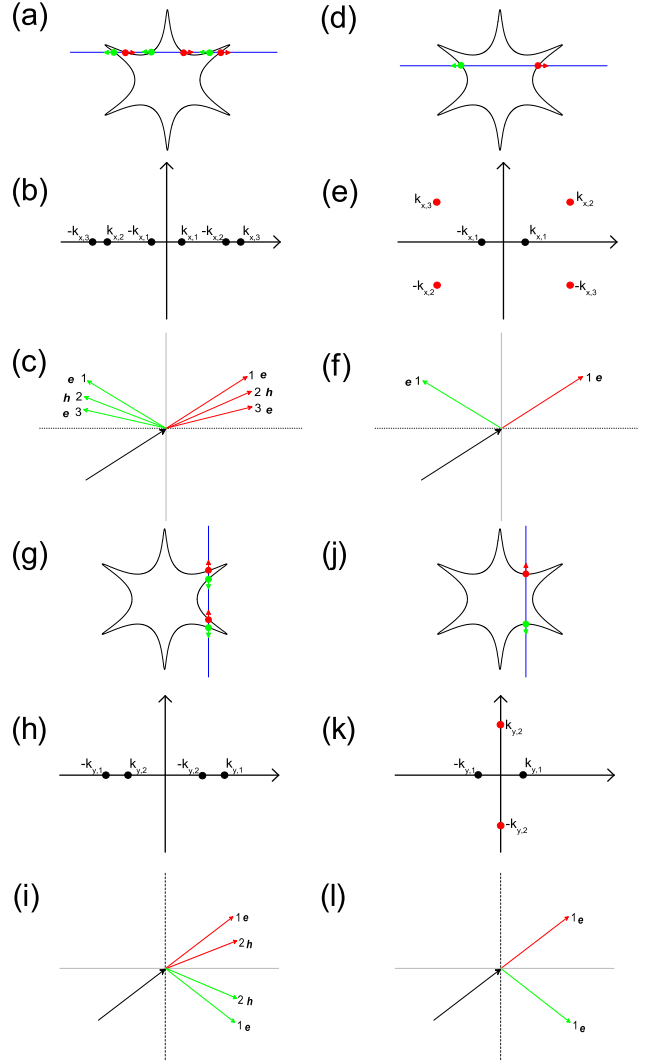}
\caption{\label{fig2} (a), (d) ((g), (j)): Two representative scattering processes when
an electron is incoming and then reflected or transmitted from the step defect
which runs along $\mathbf{y}$ ($\mathbf{x}$), where energy $E$ and momentum $k_{y}$ ($k_{x}$) are
preserved in each scattering process. The corresponding six (four) $k_{x}$ ($k_{y}$) solutions in the $k_{x} (k_{y})$ complex plane for the scattering states are displayed schematically in (b), (e) ((h), (k)).
Only the propagating reflected and transmitted waves for each situation are displayed
schematically in the (c), (f) ((i), (l)).}
\end{figure}

For type $\mathbf{II}$ solution, $k_{x,1}$ is a real number while the other four roots with relation $k_{x,2}=-(k_{x,3})^{*}$ are all complex ones, symmetrically distributed  in the $k_{x}$ complex plane, as demonstrated in Fig.2(d) and Fig.2(e), since according to Eq.(3), a complex root $k_{x}$ implies $-k_{x}$ and $\pm k_{x}^{*}$ are also complex roots. This kind of solution is always possible regardless of energy $E$. An incoming electron wave from the left will have one propagating reflected wave and only one propagating transmitted wave (see Fig. 2(f)), similar to the case appearing in most of conventional metals. However, the presence of the complex wave vectors $\pm k_{x,2}$, $\pm k_{x,3}$ means that there still exist two decaying modes on both sides of the step defect which attenuate quickly when away from the step with the same attenuation length $\mid \mathrm{Im}k_{x,2}\mid^{-1}$. Here $k_{x,1}$, $k_{x,2}$, $k_{x,3}$ are chosen as $k_{x,1}>0$, $\mathrm{Im}(k_{x,2})=\mathrm{Im}(k_{x,3})>0$ (Fig.2(e)).

Taking these considerations into account and denoting $\widetilde{\psi}_{\mathbf{k}}(\mathbf{r})=\frac{1}{\sqrt{|v_{x}(\mathbf{k})}|}\psi_{\mathbf{k}}(\mathbf{r})$ , $\mathbf{k}^{s\alpha}=(sk_{x,\alpha},k_{y})$, with $s=\pm$, one can write down the following expressions for the wave function of a generic scattering state with incident momentum $\textbf{k}=\mathbf{k}^{s\alpha}$:
\begin{eqnarray}
\nonumber \Psi_{\mathbf{k}}^{I}(\mathbf{r})&=&\widetilde{\psi}_{\mathbf{k}}(\mathbf{r})+\sum_{\beta=1}^{3}r_{\alpha\beta}^{s}\widetilde{\psi}_{\mathbf{k}^{-s\beta}}(\mathbf{r}), \ \ xv_{x}(\mathbf{k})<0 \\
\Psi_{\mathbf{k}}^{II}(\mathbf{r})&=&\sum_{\beta=1}^{3}t_{\alpha\beta}^{s}\widetilde{\psi}_{\mathbf{k}^{s\beta}}(\mathbf{r}),  \ \ xv_{x}(\mathbf{k})>0,
\end{eqnarray}%
where $r_{\alpha\beta}^{s}$ and $t_{\alpha\beta}^{s}$ ($s=\pm$, $\alpha,\beta=1,2,3$) are the reflection and transmission amplitudes for the scattering process. Since an incident electron must be a propagating plane wave, $\mathbf{k}$ has to be real, indicating that $\textbf{k}^{s\alpha}$ can be any one of the six momenta $\textbf{k}^{\pm1}$, $\textbf{k}^{\pm2}$, $\textbf{k}^{\pm3}$ for the first case, whereas $\textbf{k}$ can be just one of the two momenta $\textbf{k}^{\pm1}$ for the second case. Therefore, for type $\mathbf{I}$ solution, on the basis of $(\mathbf{k}^{+1},\mathbf{k}^{+2},\mathbf{k}^{+3},\mathbf{k}^{-1},\mathbf{k}^{-2},\mathbf{k}^{-3})$, one can construct a $6\times6$ scattering matrix $\mathbf{S}$ as follows
\begin{eqnarray}
\mathbf{S}=(\begin{array}{cc}
              t^{+} & r^{+} \\
              r^{-} & t^{-}
            \end{array}
)
\end{eqnarray}%
where $3\times3$ matrices $t^{s}$, $r^{s}$ ($s=\pm$) are defined as $t^{s}=(t^{s}_{\alpha\beta})$, $r^{s}=(r^{s}_{\alpha\beta})$ ($\alpha,\beta=1,2,3$). For type $\mathbf{II}$ solution, a $2\times2$ scattering matrix $\mathbf{S}$ on the basis of $(\mathbf{k}^{+1},\mathbf{k}^{-1})$ can be then constructed accordingly,
\begin{eqnarray}
\mathbf{S}=(\begin{array}{cc}
              t_{11}^{+} & r_{11}^{+} \\
              r_{11}^{-} & t_{11}^{-}
            \end{array}
)
\end{eqnarray}%

Since the current operator $\widehat{v}_{x}$ can be expressed as $\widehat{v}_{x}=\partial H/\partial p_{x}=p_{x}/m^{*}+v\sigma_{y}+3\lambda (p_{x}^{2}-p_{y}^{2})\sigma_{z}$, it can be checked that $<\widetilde{\psi}_{\mathbf{k}}\mid\widehat{v}_{x}\mid\widetilde{\psi}_{\mathbf{k}}>=\frac{v_{x}(\mathbf{k})}{|v_{x}(\mathbf{k})|}$. So the introduction of the factor $\frac{1}{\sqrt{|v_{x}(\mathbf{k}})|}$ into Eq.(4) is to make the expressions for the current conservation more physical and apparent, resulting in that $\mathbf{S}$ is a unitary matrix satisfying,
\begin{eqnarray}
\mathbf{S}\mathbf{S}^{+}=\mathbf{S}^{+}\mathbf{S}=\mathbf{1}.
\end{eqnarray}%

Let $\mathbf{k}=\mathbf{k}^{+1}$, we have to determine the six parameters $r_{\beta}\equiv r_{1\beta}^{+}$ and $t_{\beta}\equiv t_{1\beta}^{+}$ for $\beta=1,2,3$, so we need six equations or three spinor equations as the boundary conditions. Thanks to the fact that the Schr\"{o}dinger's equation $\{\ H(-i\partial_{x},k_{y})+U\delta(x) \}\ \Psi_{\mathbf{k}}(\mathbf{r})=E\Psi_{\mathbf{k}}(\mathbf{r})$ corresponding to Hamiltonian (1) is a third-order partial differential equation with respect to $x$, due to the warping term, we have the following boundary conditions:

\begin{eqnarray}
\nonumber \Psi_{\mathbf{k}}^{II}|_{\mathbf{r}=(0,y)}&=&\Psi_{\mathbf{k}}^{I}|_{\mathbf{r}=(0,y)},\  \partial_{x}\Psi_{\mathbf{k}}^{II}|_{\mathbf{r}=(0,y)}=\partial_{x}\Psi_{\mathbf{k}}^{I}|_{\mathbf{r}=(0,y)}\\
a^{2}\sigma_{z}(\partial_{x}^{2}\Psi_{\mathbf{k}}^{II}&-&\partial_{x}^{2}\Psi_{\mathbf{k}}^{I})|_{\mathbf{r}=(0,y)}=i\eta\Psi_{\mathbf{k}}^{II}|_{\mathbf{r}=(0,y)},
\end{eqnarray}%
which determine exactly the six parameters, with the \emph{dimensionless} ratio $\eta$ defined by $\eta=\frac{U}{v}$. The above 1st and 2nd conditions are actually the continuities of the scattering wave function and its 1st derivative at the defect line, respectively. The third condition is about discontinuity of the 2nd derivative of the wave function due to the delta-function potential, derived by integrating the Schr\"{o}dinger's equation between $x=0^{-}$ and $x=0^{+}$.

When these scattering states are fully determined by the numerical calculations, by taking into account both contributions of the reflected waves from the same side and transmitted waves from the other side, one can obtain the symmetric LDOS near the step defect by summing over all possible independent scattering states,

\begin{eqnarray} \nonumber
&&\rho(E,x;x>0) \\
&&\nonumber =\sum_{s=\pm}\int_{v_{x}(\mathbf{k})<0}\frac{d^{2}\mathbf{k}}{4\pi^{2}N_{\mathbf{k}}}|\Psi_{\mathbf{k}}^{I}(\mathbf{r})|^{2}\delta(E-E_{\mathbf{k}}^{s})\\ \nonumber
&&+\sum_{s=\pm}\int_{v_{x}(\mathbf{k})>0} \frac{d^{2}\mathbf{k}}{4\pi^{2}N_{\mathbf{k}}}|\Psi_{\mathbf{k}}^{II}(\mathbf{r})|^{2} \delta(E-E_{\mathbf{k}}^{s})\\
&&\nonumber = \sum_{s=\pm}\int\frac{d^{2}\mathbf{k}}{8\pi^{2}N_{\mathbf{k}}} \{\ |\Psi_{\mathbf{k}}^{I}(\mathbf{r})|^{2} + |\Psi_{\mathbf{k}}^{II}(\mathbf{r}_{r})|^{2} \}\ \\
&&\ \ \ \ \ \ \ \   \times\delta(E-E_{\mathbf{k}}^{s}) \\
&&=\oint_{EEC}\frac{d\emph{k}_{y}}{8\pi^{2}N_{\mathbf{k}}v_{x}(\mathbf{k})} \{\ |\Psi_{\mathbf{k}}^{I}(\mathbf{r})|^{2} + |\Psi_{\mathbf{k}}^{II}(\mathbf{r}_{r})|^{2} \}\,
\end{eqnarray}
where $\mathbf{r}_{r}=(-x,y)$. The reflection symmetry of the wave functions $\Psi_{\mathbf{k}_{r}}^{II}(\mathbf{r})=\Psi_{\mathbf{k}}^{II}(\mathbf{r}_{r})$, and $v_{x}(\mathbf{k}_{r})=-v_{x}(\mathbf{k})$ have been used in the derivation of the above equations. Here $N_{\mathbf{k}}$ is the normalization factor for the scattering state, which is equal to $2/|v_{x}(\mathbf{k})|$ if $k_{x}$ belongs to type $\mathbf{II}$ solution, but given by $1/|v_{x}(\mathbf{k})|+\sum_{\alpha=1}^{3} \{\ |r_{\alpha}|^{2}+|t_{\alpha}|^{2} \}\ /|v_{x}(\textbf{k}^{+\alpha})|$ otherwise, with $\mathbf{k}^{+1}=\mathbf{k}$. We remark here that $N_{\mathbf{k}}$ is important to be included in the expression to give the correct LDOS since it is introduced to fulfill the scattering states' orthonormal relations $N_{k}^{-1}<\Psi_{\mathbf{k}}|\Psi_{\mathbf{k}^{'}}>=\delta(\mathbf{k}-\mathbf{k}^{'})$. Eq.(10) is another expression for the LDOS which is a contour integral over $k_{y}$ on the EEC. In next sections, by making use of Eq.(9) and Eq.(10), we shall do the numerical calculations on the LDOS and perform the scaling analysis of the Friedel oscillations respectively.

When the step defect is running along other direction than $\Gamma-K$ or its two equivalent directions, a straightforward discussion leads to an analogous six-root scattering mechanism, indicating this is generic feature of the scattering process of the topological surface states in the presence of the hexagonal warping effect.

\emph{Step defect along x} ($\Gamma-K$) \emph{direction}. A generically incoming electron with momentum $\mathbf{k}$ has two transmitted waves and two reflected waves, since with $k_{x}$ a good quantum number in this situation, $\frac{k^{2}}{2m^{*}}\pm E_{\mathbf{k}}=E$, i.e., Eq.(3), is a quartic equation of $k_{y}$. It gives four roots: $\pm k_{y,\alpha}$, $\alpha=1,2$ (see Figs.2(g)-(l)). In this situation, we also have two types of solutions. For type $\mathbf{I}$, $k_{y,1}$, $k_{y,2}$ are both real numbers, as shown in Fig.2(g), which occurs when energy is above another critical value $E_{c2}$($E_{c2}=\sqrt{7}/6^{3/4}E^{*}\approx0.69E^{*}$ in the case of $E_{a}=0$)\cite{LFu} at which the EEC become concave, or the EEC has inflection points fulfilling $(\partial k_{x}/\partial k_{y})_{E}=(\partial^{2}k_{x}/\partial k_{y}^{2})_{E}=0$. To have the same sign of the group velocity $v_{y}(\mathbf{k})$, $k_{y,1}$, $k_{y,2}$ are chosen as $k_{y,1}>0$, $k_{y,2}<0$. For type $\mathbf{II}$ solution, which is always possible regardless of $E$, $k_{y,1}$ is real whereas $k_{y,2}$ is complex. $k_{y,2}$ is actually purely imaginary due to the fact that any complex root $k_{y}$ of Eq.(3) implies $-k_{y}$ and $\pm k_{y}^{*}$ are also roots, but at most two complex roots can be available. This is schematically shown in Fig.2(k). Correspondingly the reflected and transmitted propagating waves are displayed in Fig.2(i) and Fig.2(l) respectively for these two cases. Here $k_{y,1}$, $k_{y,2}$ are chosen as $k_{y,1}>0$, $\texttt{Im}k_{y,2}>0$. The exception is when $k_{x}=0$. Instead of a quartic equation, Eq.(3) in this particular case is a quadratic equation, which has only two real roots:$\pm k_{y}$. The scattering state at this special direction is free of warping parameter and so is unaffected by the warping effect.

For an incident electron with momentum $\mathbf{k}=\textbf{k}^{s\alpha}$ ($s=\pm, \alpha=1,2$), the wave function for the scattering state can be given analogously,
\begin{eqnarray}
\nonumber \Psi_{\mathbf{k}}^{I}(\mathbf{r})&=&\widetilde{\psi}_{\textbf{k}}(\mathbf{r})+\sum_{\beta=1}^{2}r_{\alpha\beta}^{s}\widetilde{\psi}_{\textbf{k}^{-s\beta}}(\mathbf{r}), \ \ \ yv_{y}(\mathbf{k})<0 \\
\Psi_{\mathbf{k}}^{II}(\mathbf{r})&=&\sum_{\beta=1}^{2}t_{\alpha\beta}^{s}\widetilde{\psi}_{\mathbf{k}^{s\beta}}(\mathbf{r}), \ \ \ \ \ \ \ \ \ \ \ \ \ \ yv_{y}(\mathbf{k})>0,
\end{eqnarray}%
with $\widetilde{\psi}_{\textbf{k}}(\mathbf{r})=\frac{1}{\sqrt{|v_{y}(\mathbf{k})}|}\psi_{\textbf{k}}(\mathbf{r})$, $\mathbf{k}^{s\alpha}=(k_{x},sk_{y,\alpha})$. Here $r_{\alpha\beta}^{s}$ and $t_{\alpha\beta}^{s}$ ($s=\pm$, $\alpha,\beta=1,2$) are the reflection and transmission amplitudes for the scattering process. Since $\mathbf{k}$ must be real, $\textbf{k}^{s\alpha}$ can be any one of the four momenta $\textbf{k}^{\pm1}$, $\textbf{k}^{\pm2}$ for type $\mathbf{I}$ solution, while can be only one of the two momenta $\textbf{k}^{\pm1}$ for type $\mathbf{II}$ solution. For the first case, on the basis of $(\mathbf{k}^{+1},\mathbf{k}^{+2},\mathbf{k}^{-1},\mathbf{k}^{-2})$, one can construct a $4\times4$ unitary scattering matrix $\mathbf{S}$ which has exactly the same form to Eq.(5) except $t^{s}=(t_{\alpha\beta}^{s})$, $r^{s}=(r_{\alpha\beta}^{s})$ ($s=\pm, \alpha,\beta=1,2$ ) are all $2\times2$ matrices. For the second case, a $2\times2$ unitary $\textbf{S}$ matrix with the exactly same form to Eq.(6) can be constructed.

Consider $\mathbf{k}=\mathbf{k}^{+1}$, we need to determine the four parameters $r_{\beta}\equiv r_{1\beta}^{+}$ and $t_{\beta}\equiv t_{1\beta}^{+}$  ($\beta=1,2$) which satisfy the following boundary conditions:
\begin{eqnarray}
\nonumber \Psi_{\mathbf{k}}^{II}|_{\mathbf{r}=(x,0)}&=&\Psi_{\mathbf{k}}^{I}|_{\mathbf{r}=(x,0)}\\
\nonumber a(E_{a}^{'}-3k_{x}^{'}\sigma_{z})\{\ \partial_{y}\Psi_{\mathbf{k}}^{II}&-&\partial_{y}\Psi_{\mathbf{k}}^{I} \}\ |_{\mathbf{r}=(x,0)}\\
&=&\eta\Psi_{\mathbf{k}}^{II}|_{\mathbf{r}=(x,0)}.
\end{eqnarray}%
Since the corresponding Shr\"{o}dinger equation $\{\ H(k_{x},-i\partial_{y})+U\delta(y) \}\ \Psi_{\mathbf{k}}(\mathbf{r})=E\Psi_{\mathbf{k}}(\mathbf{r})$ with respect to $y$ is actually a conventional sencond-order partial differential one. Accordingly, the LDOS $\rho(E,y)$ can be shown in an analogous way to have the following form,
\begin{eqnarray}
\nonumber &\rho&(E,y;y>0) \\
\nonumber &=& \sum_{s=\pm}\int \frac{d^{2}\mathbf{k}}{8\pi^{2}N_{\mathbf{k}}} \{\ |\Psi_{\mathbf{k}}^{I}(\mathbf{r})|^{2} + |\Psi_{\mathbf{k}}^{II}(\mathbf{r}_{r})|^{2} \}\ \delta(E-E_{\mathbf{k}}^{s}) \\
&=& \oint_{EEC}\frac{d\emph{k}_{x}}{8\pi^{2}N_{\mathbf{k}}v_{y}(\mathbf{k})} \{\ |\Psi_{\mathbf{k}}^{I}(\mathbf{r})|^{2} + |\Psi_{\mathbf{k}}^{II}(\mathbf{r}_{r})|^{2} \}\
\end{eqnarray}%
where $\mathbf{r}_{r}=(x,-y)$, and the normalization factor $N_{\mathbf{k}}$ is equal to $2/|v_{y}(\mathbf{k})|$ if $k_{y}$ belongs to type $\mathbf{II}$ solution, but given by $1/|v_{y}(\mathbf{k})|+\sum_{\alpha=1}^{2} \{\ |r_{\alpha}|^{2}+|t_{\alpha}|^{2} \}\ /|v_{y}(\textbf{k}^{+\alpha})|$ otherwise, with $\mathbf{k}^{+1}=\mathbf{k}$.

For simplicity of analysis, in the following discussions, $1/m^{*}$ is assumed to be zero and the effect of the particle-hole asymmetry will be neglected unless otherwise mentioned.

\begin{figure}[!htb]
\includegraphics[width=0.5\textwidth,bb=0 0 85 115]{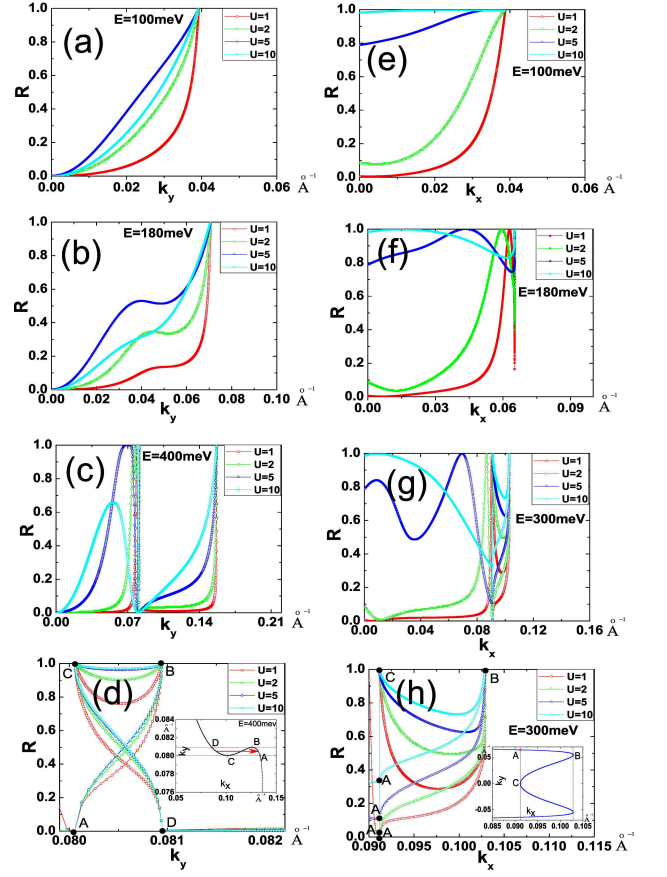}
\caption{\label{fig3} Total reflection coefficient $R$ of an incident electron as a function of the non-conserved momentum along EEC for the defect running along y($\Gamma-M$) direction for the left panels, or along x($\Gamma-K$) direction for the right panels, at different energies. (d) and (h) are the blowups of the oscillating parts of E=400meV curves in (c) and E=300meV ones in (g), respectively. Here $U$ is measured in unit of $\mathrm{eV}\cdot{{\mathrm{\AA}}}$. The insets in (d) and (h) are the relevant parts of the EECs, where the possible critical transmission points are denoted by solid dots. }
\end{figure}

\section{Transmission properties}
For \emph{defect along y} ($\Gamma-\mathrm{M}$) \emph{direction}, an incident electron at lower energy with momentum $\mathbf{k}$ has only one transmitted wave propagating along the same direction, and only one reflected wave propagating along $\mathbf{k}_{r}=(-k_{x},k_{y})$, with the other four non-propagating decaying modes localized on both sides of the defect(Figs.2(d)-2(e)). At higher energy $E>E_{c1}$, the scattering states with three transmitted and three reflected propagating modes are available(Figs.2(a)-2(b)). Since the step defect is nonmagnetic and so preserves time reversal symmetry, the backscattering between $\mathbf{k_{r}}$ and $\mathbf{k}$ if $\mathbf{k_{r}=-k}$ is forbidden. In the left panels of Fig.3 we show the total reflection coefficient $R=\sum_{\alpha=1,\mathrm{if}\   k_{x,\alpha}\ \mathrm{is}\ \mathrm{real}}^{3}|r_{\alpha}|^{2}$ of an incident electron, which clearly exhibits the absence of backscattering at $k_{y}=0$. Total reflection $R=1$ at the extreme point $k_{x}=0$ is trivial since this just means that an incident wave propagating along the step will keep propagating without being scattered. Between the backscattering point and the trivial point, $R$ changes continuously if the incident energy $E$ is so low that there is no other extreme points on the EEC (Figs.3(a)-3(b)). However at higher energy($E>E_{c1}$) with large warping effect where type $\mathbf{I}$ solutions of the scattering states are possible for Eq.(3), additional extreme points appear (see the points A, B, C, D in the inset of Fig.3(d)), and $R$ take exact values $1$ or $0$ at these points (see Figs.3(c)-3(d)). This indicates that an incident electron with momentum at points A or D will be perfectly transmitted while with momentum at points B and C will be totally reflected. The total refections at points B and C are essentially trivial, similar to the trivial point at $k_{y}=0$, since at these two extreme points, $v_{x}=(\partial E/\partial k_{x})_{k_{y}}=-v_{y}(\partial k_{y}/\partial k_{x})_{E}=0$. The existence of these resonant extreme points leads to the oscillations behavior of $R$ when $\mathbf{k}$ going through these points along the EEC (Figs.3(c)).

In the right panels of Fig.3, the total reflection coefficient $R$ for \emph{defect along x} ($\Gamma-\mathrm{K}$) \emph{direction} is exhibited. At strong scattering strength, an incident electron may be totally reflected at a resonant momentum, which increases with $U$. Moreover, the incident electrons with momentum at extreme points B and C will still be totally reflected. However, different from the former case of perfect transmission, electrons incident at point A have finite probability of being reflected, which also increases with $U$(Fig.3(h)). Unusually, the reflection coefficient $R$ is not vanishing when $k_{x}$ is approaching $0$ (Figs.3(e)-3(g)), seemingly contradicting time reversal symmetry. In fact, this is because, as mentioned before, $k_{x}=0$ is a very special direction. For any finite $k_{x}$, Eq.(3) which determines the solution of $k_{y}$ is quartic, while at exactly $k_{x}=0$, it is quadratic. This discontinuity at $k_{x}=0$ indicates that the absence of backscattering at exact $k_{x}=0$ may not be true any more at $k_{x}=0\pm$. According to Eq.(2), the equation which determines $k_{y}$ when $k_{x}\rightarrow0$ can be given by,
\begin{eqnarray}
9\lambda^{2}k_{x}^{2}k_{y}^{4}+v^{2}k_{y}^{2}-E^{2}=0.
\end{eqnarray}%
It has four roots: $k_{y}^{2}=\frac{1}{18\lambda^{2}k_{x}^{2}}\{-v^{2}\pm \sqrt{v^{4}+36\lambda^{2}k_{x}^{2}E^{2}}\}$, including two real ones, $\pm \frac{E}{v}$, as well as two complex ones: $\pm \frac{iv}{3k_{x}\lambda}$.

At exactly $k_{x}=0$, which corresponds to normal incidence, only the two real roots are available. The scattering process is actually described by a simplified Hamiltonian $H=-vp_{y}\sigma_{x}+U\delta(y)$ under the following boundary condition,
\begin{eqnarray}
(\Psi_{\mathbf{k}}^{II}-\Psi_{\mathbf{k}}^{I})|_{\mathbf{r}=(x,0)}=i\frac{\eta}{2}\sigma_{x}(\Psi_{\mathbf{k}}^{I}+\Psi_{\mathbf{k}}^{II})|_{\mathbf{r}=(x,0)},
\end{eqnarray}%
where $\eta=U/v$. Here the scattering wave functions $\Psi_{\mathbf{k}}^{I}$ and $\Psi_{\mathbf{k}}^{II}$ are expressed as,
\begin{eqnarray}
\nonumber \Psi_{\mathbf{k}}^{I}=e^{ik_{y}y}\left(
                                             \begin{array}{c}
                                               1 \\
                                               -1 \\
                                             \end{array}
                                           \right)+r_{1}e^{-ik_{y}y}\left(
                                                                     \begin{array}{c}
                                                                       1 \\
                                                                       1 \\
                                                                     \end{array}
                                                                   \right), y<0 \\
\Psi_{\mathbf{k}}^{II}=t_{1}e^{ik_{y}y}\left(
                                         \begin{array}{c}
                                           1 \\
                                           -1 \\
                                         \end{array}
                                       \right), y>0 .
\end{eqnarray}%
This gives $r_{1}=0$, $t_{1}=e^{-i\phi}$, with $\phi=2tan^{-1}(\eta/2)$, indicating that the absence of backscattering is protected by time-reversal symmetry. This is in good agreement with the well studied problem in Refs[14,17], where the delta-function potential is treated as a rectangular potential barrier in the limit of zero barrier width. The only consequence of the delta-function step potential is the discontinuity of the scattering wave function at the step line, which produces a $U$ dependent phase shift between the transmitted and incident waves.

On the other hand, when $k_{x}$ is approaching but not equal to $0$, an nearly normal incident electron with small finite $k_{x}$ has always two decaying modes on both sides of the defect, whose attenuation length $1/\kappa\simeq\frac{3\lambda k_{x}}{v}$ becomes infinitesimal small when $k_{x}\rightarrow0$. These two decaying modes make the first derivative of the scattering wave function discontinuous at the defect line. More precisely, an incident electron with momentum $\mathbf{k}=(0\pm,k_{y})$ is spin polarized along x direction, while the corresponding two decaying modes are spin polarized along y direction and are nearly exactly localized at the step edge (Fig.4(b)). Specifically, the scattering wave function for a nearly normal incident electron($k_{x}\rightarrow0$) can be thus written as,

\begin{figure}[!htb]
\includegraphics[width=0.5\textwidth,bb=0 0 245 235]{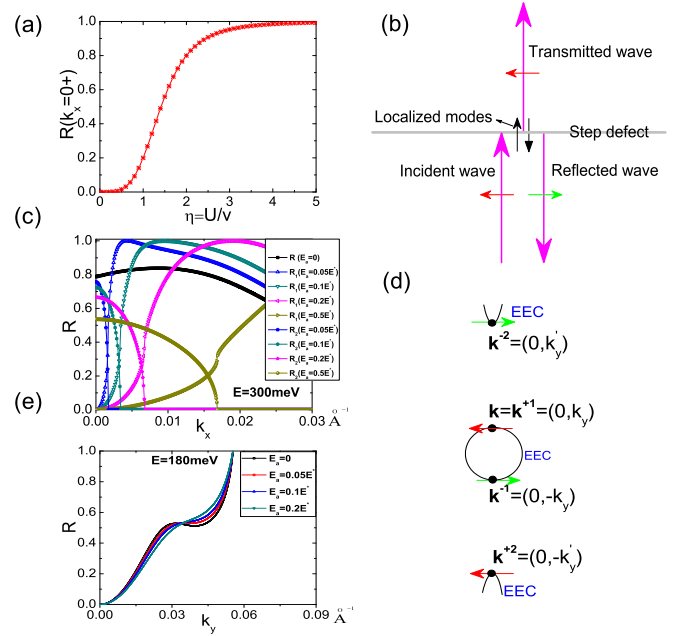}
\caption{\label{fig4} (a)Reflection coefficient $R$ as a function of the dimensionless scattering strength $\eta=U/v$ for a nearly normal incident electron, when the \emph{defect runs along x direction}. The relevant scattering state is schematically shown in (b). In the presence of particle-hole asymmetric term, a new additional EEC occurs near y axis at large momentum. Reflection coefficients $R_{1}$ (reflection to the same EEC), $R_{2}$(reflection to the other EEC) on a $E=300meV$ EEC are shown in (c) for $U=5\mathrm{eV}\cdot{{\mathrm{\AA}}}$, while correspondingly the scattering state is schematically exhibited in (d). (e)Reflection coefficient $R$ on $E=180meV$ EEC for different particle-hole asymmetric energies $E_{a}$, when the \emph{defect runs along y direction}. }
\end{figure}

\begin{eqnarray}
\nonumber \Psi_{\mathbf{k}}^{I}=e^{ik_{y}y}\left(
                                             \begin{array}{c}
                                               1 \\
                                               -1 \\
                                             \end{array}
                                           \right)+r_{1}e^{-ik_{y}y}\left(
                                                                     \begin{array}{c}
                                                                       1 \\
                                                                       1 \\
                                                                     \end{array}
                                                                   \right) \\
\nonumber                                                                   +r_{2}e^{\kappa y}\left(
                                                                                              \begin{array}{c}
                                                                                                1 \\
                                                                                                i \\
                                                                                              \end{array}
                                                                                            \right), y<0 \\
\Psi_{\mathbf{k}}^{II}=t_{1}e^{ik_{y}y}\left(
                                         \begin{array}{c}
                                           1 \\
                                           -1 \\
                                         \end{array}
                                       \right)+t_{2}e^{-\kappa y}\left(
                                                                   \begin{array}{c}
                                                                     1 \\
                                                                     -i \\
                                                                   \end{array}
                                                                 \right), y>0 ,
\end{eqnarray}%
where $\kappa=\frac{v}{3\lambda k_{x}}$. According to the boundary conditions of Eq.(12), the coefficients can be easily solved to give $r_{1}=\frac{-i\eta^{2}/2}{1-\eta^{2}/2+i\eta}$, $t_{1}=\frac{1}{1-\eta^{2}/2+i\eta}$, $r_{2}=\frac{(i+1)(\eta^{2}-i\eta)/2}{1-\eta^{2}/2+i\eta}$, $t_{2}=\frac{(i+1)\eta/2}{1-\eta^{2}/2+i\eta}$, giving rise to the final result of the reflection coefficient,
\begin{eqnarray}
R(k_{x})=\left\{
\begin{array}{c}
      0, \ \ \ \ \ \ \ \ \  k_{x}=0 \\
      \frac{\eta^{4}/4}{1+\eta^{4}/4}, \ \ k_{x}\rightarrow0
    \end{array}
\right.
\end{eqnarray}%
This implies the discontinuity of $R$ at $k_{x}=0$. We remark here that although this phenomenon is caused by the warping effect, the reflection coefficient $R(k_{x}\rightarrow0)$ is independent of the warping parameter, which means taht the warping parameter is relevant in this particular case, even at lower energies. This is a unique feature of a delta-function potential. As a comparison, for a step-function potential,\cite{ZhuBF,HuJP} one can construct an analogous scattering wave function to obtain $r_{2}=t_{2}=r_{1}=0$, $t_{1}=1$ due to the continuity of the first derivative of the scattering wave function at the step line, indicating the continuity of the reflection coefficient($R_{1}(k_{x}\rightarrow0)=R_{1}(k_{x}=0)=0$ and the absence of decaying modes.

Since spin is not conserved in the scattering process, the localized spin-polarized modes act as a "magnetic barrier" for the propagating waves and the existence of them makes the nonmagnetic defect "magnetic", which leads to the scattering between the two nearly time-reversal related states. We explain this as follows. Consider a \emph{nearly normal incident massless Dirac fermion without warping effect} on a magnetic step defect, whose scattering potential is described by $(U+\mathbf{V}_{m}\cdot\bm{\sigma})\delta(y)$. The scattering wave function is described by the exactly same equation given by Eq.(16), but under a slightly different boundary condition to Eq.(15) with the replacement of $\eta$ in (15) by $\eta+\frac{1}{v}\mathbf{V}_{m}\cdot\bm{\sigma}$. It can be checked that if $\mathbf{V}_{m}$ is chosen as $\mathbf{V}_{m}/v=\frac{\eta^{2}}{8-2\eta^{2}+\eta^{4}}(-2\eta,-\eta^{3},4)$, the exactly same results for $r_{1}$ and $t_{1}$ described by Eq.(16) can be obtained. This gives an equivalent physical picture that for nearly normal incidence the localized modes can be seen as a delta-function magnetic barrier positioned at the step defect for the propagating waves. This naturally explains the phenomenon of finite reflection.

In more realistic situation, this problem of "presence of backscattering" can be solved if a small particle-hole asymmetric term is taken into account in Hamiltonian (1)(Fig.4(c)). The physical picture is that: If $E_{a}\neq 0$, there exists an additional segmental EEC around y axis at relatively large momenta(Fig.4(d)). Instead of the two decaying modes in the case of $E_{a}=0$, there exist two propagating modes at large momenta with spin polarization along x axis rather than y axis. Therefore the scattering wave function for a nearly normal incident electron can be given by,
\begin{eqnarray}
\nonumber \Psi_{\mathbf{k}}^{I}=e^{ik_{y}y}\left(
                                             \begin{array}{c}
                                               1 \\
                                               -1 \\
                                             \end{array}
                                           \right)+r_{1}e^{-ik_{y}y}\left(
                                                                     \begin{array}{c}
                                                                       1 \\
                                                                       1 \\
                                                                     \end{array}
                                                                   \right) \\
\nonumber                                                                   +r_{2}e^{-ik_{y}^{'} y}\left(
                                                                                              \begin{array}{c}
                                                                                                1 \\
                                                                                                -1\\
                                                                                              \end{array}
                                                                                            \right), y<0 \\
\Psi_{\mathbf{k}}^{II}=t_{1}e^{ik_{y}y}\left(
                                         \begin{array}{c}
                                           1 \\
                                           -1 \\
                                         \end{array}
                                       \right)+t_{2}e^{ik_{y}^{'}y}\left(
                                                                   \begin{array}{c}
                                                                     1 \\
                                                                     1 \\
                                                                   \end{array}
                                                                 \right), y>0 ,
\end{eqnarray}%
where $k_{y}^{'}=k_{y}+(E_{a}^{'}a)^{-1}$. Analogously, the coefficients can be determined to obtain $r_{1}=t_{2}=0$, $r_{2}=\frac{\eta}{-\eta+i\sqrt{1+4E_{a}^{'}E^{'}}}$, $t_{1}=1+r_{2}=\frac{i\sqrt{1+4E_{a}^{'}E^{'}}}{-\eta+i\sqrt{1+4E_{a}^{'}E^{'}}}$, where $\eta=U/v$. This gives rise to the reflection coefficients $R_{1}(k_{x}\rightarrow0)=|r_{1}|^{2}=0$, $R_{2}(k_{x}\rightarrow0)=|r_{2}|^{2}=\frac{\eta^{2}}{\eta^{2}+1+4E_{a}^{'}E^{'}}$. This means a nearly normal incident electron has no backscattering with $R$ continuously reduced to zero when $k_{x}\rightarrow0$, but still has finite probability to be reflected to one segment of the large-momentum EEC, with the probability $R_{2}$ increasing as a function of the scattering strength $U$.

While whether or not there is particle-hole asymmetry dramatically affects the situation where the defect runs along x axis, it actually hardly affects the other situation where the defect runs along y axis. As an example, at different small particle-hole asymmetric energies $E_{a}$, Fig.4(e) shows the reflection coefficient $R$ on $E=180$meV EEC, which is nearly unchanged with $E_{a}$. This feature is expected since the inclusion of a small particle-hole asymmetric term only shifts slightly the values of the roots of each solution, but doesn't change the types of the solutions for any scattering states.

\begin{figure}[!htb]
\includegraphics[width=0.5\textwidth,bb=0 15 98 150]{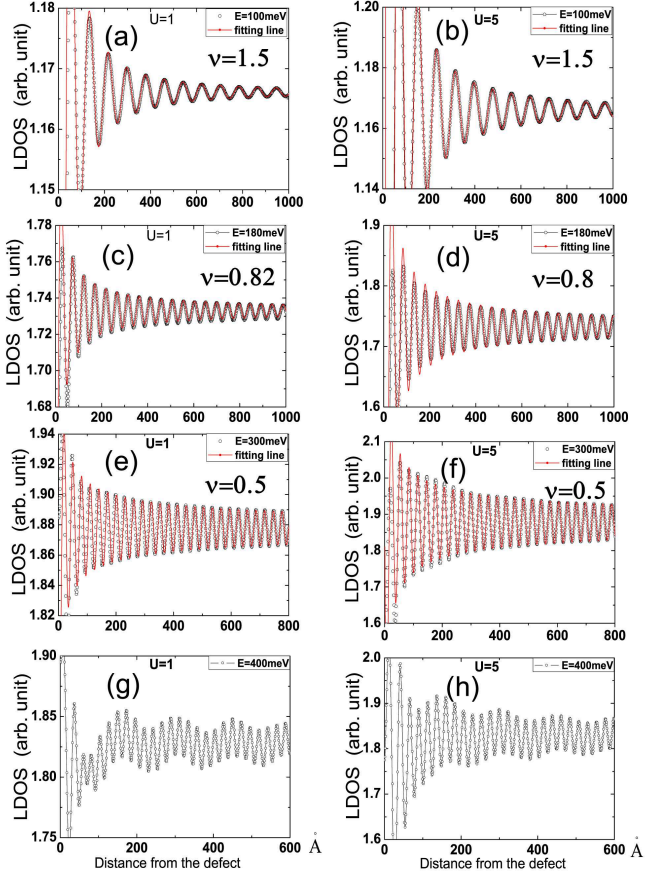}
\caption{\label{fig5} LDOS as a function of the distance from the step
for $U=1\mathrm{eV}\cdot{{\mathrm{\AA}}}$ and $U=5\mathrm{eV}\cdot{{\mathrm{\AA}}}$ at different energies E. The defect runs along y($\Gamma-M$) direction. The (red) solid lines are the fitting power-law decay sinusoidal functions $\propto \frac{sin(\Delta kx+\varphi)}{x^{\nu}}$, with $\nu$ the decay indices.}
\end{figure}

\section{LDOS and FTLDOS}
The electrons' interference patterns due to the scattering between states on the EEC can be reflected by the LDOS in the vicinity of step defect. Although there exist decaying modes for most of the scattering states, the asymptotic behavior of the Friedel-like oscillations of the LDOS is believed to be mainly determined by the interferences between the propagating waves. In most cases at low energies, the dominant contribution is coming from the interference between the incident and reflected propagating waves, whereas at high enough energies, further contributions from the interferences between the transmitted waves becomes possible and competing, leading to multi-periodic LDOS oscillations. At lower energy, the numerical LDOS exhibits the behavior of power-law decay at long distance from the defect. To explore this asymptotic behavior, we assume the long-distance behavior is dominated by the interference between the incident and reflected propagating waves and follow the scaling analysis of Ref\cite{Biswas,QinLiu} on the LDOS oscillations, since other possibilities can be discussed straightforwardly in an analogous way. This gives the following asymptotic behavior of the LDOS: $\delta\rho(x,E)\propto \frac{sin(\Delta kx+\phi)}{x^{\nu}}$, with the power-law decay index $\nu=(1+\alpha+\beta)/\gamma$. Here $\alpha$, $\beta$ and $\gamma$ are the power exponents of the expansions around the extreme points for the quantities: reflection amplitude $r$, overlap of the spinor wave functions of incident and reflected propagating waves $\chi_{\mathbf{k}}^{(+)+}\chi_{\mathbf{k}}^{(+)}$, and the momentum transfer $\Delta k_{x}$, where near the extreme points $r\propto \delta k_{y}^{\alpha}$, $\chi_{\mathbf{k}}^{(+)+}\chi_{\mathbf{k}}^{(+)}\propto \delta k_{y}^{\beta}$, $\Delta k_{x} \propto \Delta k+\delta k_{y}^{\gamma}$.

In Fig.5 we show the Friedel-like oscillations of LDOS $\rho(E,x)$ at different representative energies for \emph{step defect along y direction}. At lower energy $E<340$meV, the quasi-period of LDOS oscillations is consistent with the inverse of the wave vectors connected by the corresponding pair of the two extreme points on EEC (see the pairs of red circles in Fig.1), whereas at higher energy $E>340$meV, multi-periodic oscillations become obvious owing to the competition of the multiple scattering processes on the EEC(see Figs.5(g), and 5(h)). The decay behavior of the LDOS demonstrates that the oscillations decay much more quickly at $E=100$meV where the EEC has a circle shape than at higher energy where the EEC has a shape of hexagon or hexagram, in agreement with STM experiments.\cite{ZhuBF,STM} The asymptotic behavior of the LDOS oscillations at $E=100$meV can be very well fitted by a $x^{-3/2}$ power-law decay sinusoidal function. This is in agreement with the scaling analysis, since $\alpha=\beta=1$ i.e., both the reflection coefficient and the spinor overlap are zero due to the absence of backscattering between the pair of the extreme points ($\mathbf{q}_{1}$ in Fig.1), as well as $\gamma=2$, giving rise to $\nu=3/2$.  At energy as high as $E=300$meV, the LDOS oscillations can be well fitted by a $x^{-1/2}$ power-law decay function(see Figs.5(e) and 5(f)). This is interpreted as that the system at this energy is dominated by a scattering process between the two extreme points connected by a vector $\mathbf{q}_{2}$ as shown in Fig.1, where $\alpha=\beta=0$ and $\gamma=2$. This is analogous to the scattering processes occurring in conventional metals, since the interference scattering occurs between the points which have both finite refection and finite spinor overlap. At $E=180$meV which is intermediate between the above two energies, the LDOS can be better fitted by a power-law decaying function with decay index $\nu\approx0.82$. This can be understood since the EEC at $E=180$meV is nearly nesting. The existence of both $\mathbf{q}_{1}$-like and $\mathbf{q}_{2}$-like scattering processes and their competition lead to a power-law decay of LDOS with decay index $\nu$ between $3/2$ and $1/2$. At high enough energies such as $E=400$meV (see Figs.5(g) and 5(h)), the LDOS oscillations are dominated by multiple scattering processes so that the asymptotic behavior exhibits beat feature which can be hardly fitted by any power-law decay functions.

\begin{figure}[!htb]
\includegraphics[width=0.5\textwidth,bb=0 18 110 140]{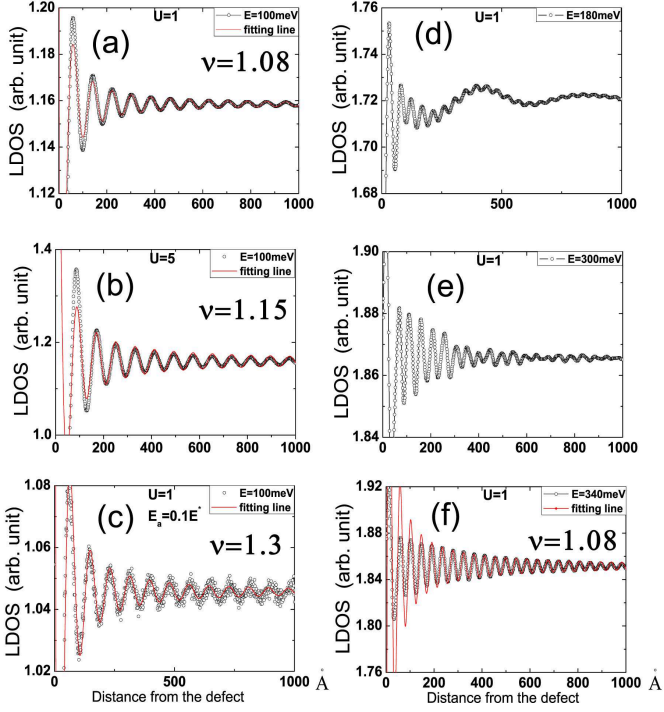}
\caption{\label{fig6} LDOS as a function of the distance from the step
for (a)U=1, $E_{a}=0$,(b)U=5, $E_{a}=0$, and (c)U=1, $E_{a}=0.1E^{*}$ at E=100meV, while (d),(e),(f) are the LDOS for U=1, $E_{a}=0$ at E=180meV, 300meV, 340meV respectively, where the defect runs along x ($\Gamma-K$) direction. The (red) solid lines are the fitting power-law decay sinusoidal functions $\propto \frac{sin(\Delta ky+\varphi)}{y^{\nu}}$, with $\nu$ the decay indices. Here $U$ is measured in unit of $\mathrm{eV}\cdot{{\mathrm{\AA}}}$.}
\end{figure}

Accordingly the LDOS $\rho(E,y)$ for \emph{step defect along x direction} is exhibited in Fig.6. The LDOS oscillations can be fitted by a power-law function only at relatively lower energies or at energies around $E=E_{c1}$ whose EEC has nesting segments. At $E=100$meV, the fitting decay index is about $1.08$(Fig.6(a)), quite different from the former case at the same energy whose fitting decay index $\nu=1/2$. This is surprising because  at this energy, the EEC is nearly a circle and the warpping effect is small, which seemingly leads to the conclusion that the transport  along y direction should be similar to that along x direction. This behavior could be understood if we note that there exists finite "backscattering" near the two extreme points related by time reversal symmetry, so $\alpha$ should be $0$ instead of $1$. $\beta$ is still equal to $0$ since the two spinors are still orthogonal to each other. Therefore $\nu=1$ is expected for the low-energy asymptotic behavior of the LDOS. Taking a small particle-hole asymmetric term into account would qualitatively change  the power-law decay behavior(see Fig.6(c)). Since the absence of backscattering is recovered by this term, the power-law decay index $\nu$ should return back to $3/2$, qualitatively consistent with our numerical result(Fig.6(c)). The appearance of small fluctuations in the LDOS with quasi-period about several angstroms in the presence of particle-hole asymmetry is a generic feature which is due to the interference scattering between the two extreme points on different EECs, one on the original EEC, the other on the additional large-momentum EEC(Fig.4(d)). At intermediate energies such as $E=180$meV(Fig.6(d)) or $E=300$meV (Fig.6(e)), the LDOS oscillations can be hardly fitted by any power-law decay functions due to the presence of  destructive interference scatterings.

\begin{figure}[!htb]
\scalebox{1.0}[1.0] {\includegraphics[6,47][250,310]{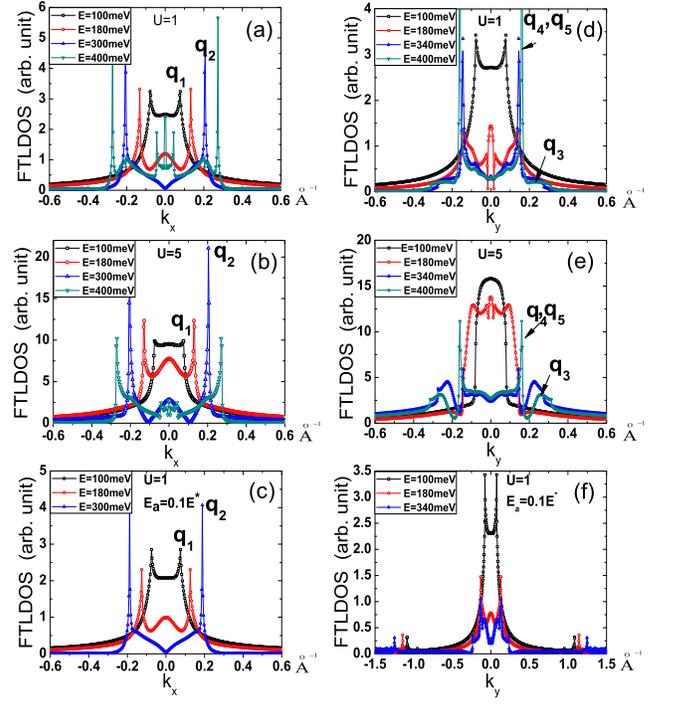}}
\caption{\label{fig7} The magnitudes of the Fourier transforms of the LDOS at different energies where the step defect runs along y($\Gamma-M$) direction are shown on the left panels. Those along x($\Gamma-K$) direction are shown on the right panels. The parameters are chosen as U=1, $E_{a}=0$ for (a), (d), and U=5, $E_{a}=0$ for (b), (e), while U=1, $E_{a}=0.1E^{*}$ for (c), (f). Here $U$ is measured in unit of $\mathrm{eV}\cdot{{\mathrm{\AA}}}$. The peaks shown in Fig.1 which are relevant to the characteristic scattering wave vectors $\mathbf{q}_{1}-\mathbf{q}_{5}$($q_{4}\approx q_{5}$) are exhibited explicitly.}
\end{figure}
To further analyze the interference patterns revealed by the LDOS, we investigate the Fourier transforms of the LDOS (FTLDOS), $\delta\rho(E,k_{x}(k_{y}))$, whose magnitudes are shown in Fig.7. The FTLDOS has been used to explore the interference patterns of a point impurity on the surface of a topological insulator.\cite{HuJP,Franz} The FTLDOS curves here are symmetric about zero Fourier momentum so let us focus on the positive-momentum part. At lower energy with weak warping effect there is always a broad peak at $k_{x}=0(\mathrm{or}  k_{y}=0)$. This can be  interpreted as follows: For each scattering state on the low-energy EEC, there exist decaying modes residing in the vicinity of the step. Although the decaying modes for different scattering states have different attenuation lengths, they all contribute a peak positioned at $k_{x}=0$ (or $ k_{y}=0$), giving rise to the appearance of a broad peak there. Except the zero-momentum peak, at lower energy there generically exists another peak characterizing the oscillations of the LDOS, such as $\mathbf{q}_{1}$ peak in Fig. 7 at $E=100$meV. However at higher energy with large warping effect, multi-peak structure appears where one of the peaks is dominant(see E=400meV curves in Fig.7). These multiple peaks are interpreted as the competition among various scattering processes between the multiple pairs of the extreme points on the EEC. For example, when the step runs along $\Gamma-\mathrm{K}$, there are three competing characteristic scattering processes at $E=400$meV(shown in Fig.1 as the characteristic wave vectors $\mathbf{q}_{3}$, $\mathbf{q}_{4}$, $\mathbf{q}_{5}$) with $\mathbf{q}_{5}$ process dominating, since it has the lowest decay index $\nu=1/2$, while there are two competing characteristic scattering processes at $E=340$meV with $\mathbf{q}_{4}$-like process dominating owing to the absence of $\mathbf{q}_{5}$-like process, in agreement with STM experiment.\cite{STM} Interestingly, there exists an additional peak at $E=400$meV for the first situation (see Figs.7(a),7(b)) at a lower momentum $k_{x}\simeq0.04{\mathrm{\AA}}^{-1}$. This additional minor peak is recognized as a result of the interference scattering process between the two points connected by a nesting vector, shown in the inset of Fig.3(d), where the two points lie on segments AB and CD respectively. This peak is only possible when the warping effect is so large that $E>E_{c1}$. On the other hand, compared with $E_{a}=0$ case, the FTLDOS in Fig.7(c) for the case with $E_{a}=0.1E^{*}$ is nearly unchanged, while that in Fig.7(f) shows a small peak appearing at large momentum, and a sharper peak at lower energy indicating the modification of the power-law decay index. This further confirm the point that the inclusion of a small particle-hole asymmetric term has little effect on the scattering property from a defect along y axis whereas has dramatical impact on that from a defect along x axis.

\section{Summary}
In summary, we have investigated the effect of a step defect on the surface states of a topological insulator with strong warping effect in a quantum-mechanical approach. Scattering properties are found to depend strongly on the extending direction of the defect. At high energy with large warping effect, there exist critical directions along which an incident electron will exhibit perfect transmissions when the defect runs along $\Gamma-\mathrm{M}$, whereas an incident electron will generally exhibit finite reflections when the defect runs along $\Gamma-\mathrm{K}$. Specifically, for  nearly normal incidence on the defect along $\Gamma-\mathrm{K}$, finite reflection always exists, which is explained as the consequence of the existence of the localized decaying modes at defect line. The Friedel oscillations of the LDOS show power-law decay, consistent with STM experiments on $\mathrm{Bi}_{2}\mathrm{Te}_{3}$. The electrons' interference pattern of a step defect can be characterized by the Fourier transform of the LDOS, which exhibits a single peak with finite momentum at lower energy while shows multi-peak structure with one of the peaks dominant at higher energy.

\emph{Note added}. When finalizing this work we became aware of a recent related unpublished paper,\cite{Hungary} in which by making use of a similar framework of a quantum-mechanical approach, only the case for the step defect running along $\Gamma-\mathrm{K}$ is considered and an asymptotically exponential decay of the LDOS is found.

\begin{acknowledgments}
We thank D. G. Zhang for helpful discussions. This work was supported by the Texas Center for Superconductivity at the University of Houston and by the Robert A. Welch Foundation under the Grant No. E-1146. J. An was also supported by NSFC (China) Projects No. 11174126 and No. 10874073.
\end{acknowledgments}

\end{document}